# Mapping (USPTO) Patent Data using Overlays to Google Maps

Loet Leydesdorff [i] and Lutz Bornmann[ii]

**Abstract**

A technique is developed using patent information available online (at the US Patent and Trademark Office) for the generation of Google Maps. The overlays indicate both the quantity and quality of patents at the city level. This information is relevant for research questions in technology analysis, innovation studies and evolutionary economics, as well as economic geography. The resulting maps can also be relevant for technological innovation policies and R&D management, because the US market can be considered the leading market for patenting and patent competition. In addition to the maps, the routines provide quantitative data about the patents for statistical analysis. The cities on the map are colored according to the results of significance tests. The overlays are explored for the Netherlands as a "national system of innovations," and further elaborated in two cases of emerging technologies: "RNA interference" and "nanotechnology."

---

[i] University of Amsterdam, Amsterdam School of Communication Research (ASCoR), Kloveniersburgwal 48, 1012 CX Amsterdam, The Netherlands; loet@leydesdorff.net.
[ii] Max Planck Society, Administrative Headquarters, Hofgartenstrasse 8, D-80539 Munich, Germany; bornmann@gv.mpg.de.

## 1. Introduction

In a series of previous studies, we developed interactive overlays of Google Maps which show the geographical distributions of publication and citation data from the Web-of-Science or Scopus (Bornmann & Leydesdorff, 2011; Bornmann *et al.*, 2011; Leydesdorff & Persson, 2010). In this study, we generalize this approach to patent data by using (automatic) downloads from the database of the US Patent and Trademark Office (USPTO). The city-nodes on Google Maps can be colored in accordance with the percentile ranks of the citation rates of the patents in the downloaded sample and/or the distribution of patents over cities. Our objective is to develop a generic tool that can be used for both practical[1] and research purposes: the distributions of patents among nations, cities, and regions is of interest for discussions in economic geography, innovation studies, and evolutionary economics, as well as technology policy and forecasting.

## 2. Theoretical relevance

Patents can be considered as indicators of inventions. The development of an indicator, however, has no intrinsic theoretical value, but can serve for the operationalization of a variable in other domains, or as an instrument facilitating further research. In a pioneering study about the economic sources of innovative activity, for example, Schmookler (1962) used US patents to show that inventions would not lead economic activities innovatively, but co-varied or even lagged behind market developments.

---

[1] For a practical application, see at http://www.technologytransfertactics.com/content/2011/11/16/researchers-map-uspto-patent-data-using-google-maps/#more-11636 .

However, the focus of Schmookler's analysis was on the period before World War II. The issue of the measurement of invention by using, among other things, patents became urgent with the specification of the "residual factor" in the economy by Abramowitz (1956; cf. Fabricant, 1954; Solow, 1957).

The "residual factor" is defined as economic growth that cannot macro-economically be attributed to input factors such as labor and capital, and is therefore held to represent upgrading of the labor force and technological innovation; in short, the advance of socially organized knowledge (Denison, 1964: 19). Can patent statistics provide us with an instrument to measure the sources of wealth generated from R&D (Grilliches, 1984; Mansfield, 1961, 1998; Pavitt, 1984; Rosenberg, 1976)? In his Presidential Address to the American Economic Association, Grilliches (1994: 10) considered the consequential worsening of measurement problems as a data constraint on economic analysis!

When one extends the production function for explaining economic output in terms of capital and labor to the "total factor productivity" (TFP) approach, one needs an operationalization of the knowledge input into the production process (Jaffe *et al.*, 1993; Coe *et al.*, 2009). How does knowledge production in R&D spill over into economic innovation and growth (Boschma, 2005; Frenken, 2009)? At the time, Grilliches (1994: 16) pointed to the work of Jaffe *et al.* (1993) about measuring patents and patent citations as a possible direction for the solution of this measurement problem. In their groundbreaking book entitled "Patents, Citations & Innovations: A Window on the Knowledge Economy," Jaffe & Traitenberg (2002) analyzed the USPTO database using almost three million US patents granted between January 1963 and December 1999, and more than 16 million citations made to these patents between 1975 and 1999.[2]

Patents have thus become a strategic asset in both business and policy-making. In the USA, the Bayh-Dole Act of 1980 gave universities (and other nonprofit institutions, as well as small business) the right to retain the property rights to inventions deriving from federally funded research. Using patent citation relations and the citation relations between patents and publications, Narin & Olivastro (1988, 1992) studied the question of whether an increasing link between technology and public science had been fostered in the US economy (cf. Boyack & Klavans, 2008; Grupp, 1996; Verspagen, 2006). Sampat (2006) used patent citations to study the question of whether the quality of university patents—in terms of patent citations—had been improved by the Bayh-Dole Act (cf. Leydesdorff & Meyer, 2010; Sampat *et al.*, 2003). Most recently (September 16, 2011), President Obama signed into law the *Leahy-Smith America Invents Act* (H.R. 1249), a bipartisan, bicameral bill that updates the US patent system in order to encourage innovation, job creation, and economic growth.

Using the model of a Triple Helix of university-industry-government relations, one can distinguish the dynamics of knowledge production, markets, and governance as three interacting, but analytically different mechanisms that may interact synergetically in the shaping of a knowledge-based economy (Leydesdorff, 2010). In the economy, patents serve for the protection of intellectual property by generating windfalls for exclusive exploitation. In policy-making, knowledge spillovers into the regional or national economy are central. The patent database can be considered as offering a window on these complex systems because patents can be considered as output of R&D, input to the “knowledge-based” economy, and intellectual property to be retained in institutional frameworks.

[2] This data was also matched with data of all firms traded on the US stock market using Compustat. This data was made available on a CD-Rom included to the book.

When several different perspectives are analytically possible, the visualization sometimes raises further research questions. Our visualizations offer us a kind of “patent radar” that may signal unexpected information. We focus on three questions in these explorations:

1. How is patenting in the US geographically distributed in the case of the Netherlands?
2. How is patenting in the US evolving along the innovation trajectory of RNA interference?
3. How are US patents internationally distributed in the case of nanotechnology?

## 3. Patent databases

Legislation has historically been national, and accordingly patent offices have initially been organized at national levels. The database of the U.S. Patent and Trade Office (USPTO) contains all the data since 1790. Patents are retrievable from this date as image files, and after 1976 also as full text. The html-format allows us to study the patents in considerable detail (Leydesdorff, 2004).

The European Patent Office (EPO) was established as a transnational patent office in 1973. This database is also online, but in a less accessible pdf-format. Furthermore, 144 nations are currently signatories of the Patent Cooperation Treaty (PCT) of 1970, which mandated the World Intellectual Property Organization (WIPO) in Geneva to administer fee-based services (since 1978). This publicly accessible database is organized in the html-format (Leydesdorff, 2008). Like the European Union, several world regions have established regional patent offices.

The various offices provide applicants with a number of choices which imply different procedures and timelines. For example, the USPTO hitherto operates on the basis of “first to invent,” while the EPO uses “first to file” as a criterion. (As of March 2013, the USPTO will change to the “first to file” criterion under the America Invents Act of 2011). If the applicant wishes to protect an invention in countries outside the country of its origin, s/he can file for a patent in each country in which protection is desired, or to a regional office (e.g., EPO), or file an international application under the Patent Cooperation Treaty procedure. Various factors (e.g., the costs of patenting, the time taken to grant patents, differences in rules regarding the scope of patents, etc.) influence the decision on whether to follow one procedure or another.

This variety of (partially overlapping) procedures complicates the use of patent statistics. In the USPTO the inventor and his/her attorney are obliged to provide a list of references describing the state of the art (Michel & Bettels, 2001). EPO examiners, and not inventors or applicants, add the large majority of patent citations (Criscuolo & Verspagen, 2008). While in the USPTO applications are examined automatically, the EPO considers an application as a request for a "patentability search report." These reports contain citations from patents and non-patent documents that have either been suggested by the inventor or added by the patent examiner (Criscuolo, 2004: 92f.).

Applications at the WIPO for intellectual property protection under the regime of the PCT protocol require an International Search Report (ISR) and a written opinion by the examiner about the patentability of the invention (OECD, 2005: 57; 2008). Although the initial investigations are usually carried out by the receiving offices, the international extension can be expected to lead to a further streamlining of the patent citations with reference to their economic value and legal protection against possible litigation in court.

From the perspective of information science and technology, patent classifications provide us with the outcomes of major investments of the patent offices to organize the patents intellectually. The International Patent Classification (IPC) uses a 12-digit code containing 70,000 categories (WIPO, 2011). The European classification system (ECLA) builds on this IPC and extends the number of categories to approximately 135,000. The USPTO has its own classification system; this classification scheme currently employs more than 430 classes and 140,000 subclasses. However, the USPTO databases can also be searched using the IPC as search terms.

Attempts to map patent data for the purpose of analyzing economic activities in terms of the technologies involved have been moderately successful. The OECD has for this purpose defined "triadic patent families" which counteract upon "home advantage effects" in the various databases (Criscuolo, 2006). A patent is a member of a patent family *if and only if* it is filed at the European Patent Office (EPO), the Japanese Patent Office (JPO), and is granted by the US Patent & Trademark Office (USPTO) (Criscuolo, 2006). Integration of the various databases into the single framework of PATSTAT has been developed by EPO for use by (inter)governmental organisations and academic institutions. However, this database is fee-based.

Alternatively, searches at Google Patents (at http://www.google.com/patents) provide full text patents for free, but the search engine is not organized for searches to specific fields or categories. Furthermore, the Derwent Innovation Index of Thomson-Reuters should be mentioned in this context as a commercial alternative. This database is now integrated into the Web of Knowledge containing also the *Science Citation Index*. However, these patent records do not currently contain address information, and the citation information is made not machine-readable.

In addition to its transparency, the USPTO database has the advantage of being the prime indicator of new technological inventions, and therefore the most relevant one for innovation policies (Narin & Olivastro, 1988; Jaffe & Trajtenberg, 2002). The US market for patents can be considered as most competitive. However, USPTO data is organized in two systems with separate search engines for patent applications and granted patents, respectively (at http://patft.uspto.gov/). Patent applications follow the research front more closely than the granted patents, but do not contain citation information. In line with our objective we make routines available for organizing and mapping downloads from both databases.

## 4. Patent maps and interactive overlays

In the bibliometric databases such as Scopus and the Web of Science (WoS), one can consider journals and countries as the two main dimensions of organization (Small & Garfield, 1985). Analogously to journals, patent classifications can be considered as the intellectual organizers which allow us to study citation flows among categories in terms of geographically positioned units such as research laboratories, industries, or whole nations. One step further one can use concordance tables between patent classification schemes and industrial sectors for the mapping of patents in economically relevant categories (Schmoch *et al.*, 2003). This longer-term perspective of mapping and tracing innovations across borders among databases can set a research program for the information sciences, also in support of discourses in neighboring disciplines. As Grilliches (1994: 14) formulated: "Our current statistical structure is badly split, there is no central direction, and the funding is heavily politicized."

One of us attempted to map a year of data from the WIPO database comprehensively in terms of IPC classifications (Leydesdorff, 2008; cf. Boyack & Klavans, 2009). However, mapping an index in terms of co-classifications is different from mapping relational data (e.g., citations) using an index. In the meantime, several research teams have made progress in generating a basemap for interactive overlays using IPC terms for the classification (Newman *et al.*, 2011; Schoen *et al.*, 2011). In this study, however, we project the intellectual organization onto the geographical map. As noted, the study follows up on our previously developed methods to map publication and citation data in this way (Bornmann & Leydesdorff, 2011; Leydesdorff & Persson, 2010). The methodology is combined with an existing routine to retrieve data from the USPTO database for citation analysis (Leydesdorff, 2004).

Using the geographical dimension for the mapping is less complex because the baseline map does not have to be constructed from the data as in the case of a classification structure (Rafols & Leydesdorff, 2009). The technique of using overlays interactively and on top of basemaps was gradually developed during the last decade by Boyack *et al.*, (2005); Klavans & Boyack (2009); Leydesdorff & Rafols (2009; 2012); Moya-Anegón *et al.* (2004; 2007); Rafols *et al.*, (2010); Rosvall & Bergstrom (2009); and others. Overlay files enable users to select their focus of interest with reference to their specific research questions. As noted, we analyze a map of a national system—the Netherlands—and study two cases of emerging technologies—RNA interference and nanotechnology—as examples.

## 5. Methods

A set of dedicated routines was developed which can be downloaded by the user at http://www.leydesdorff.net/software/patentmaps/index.htm. This webpage also contains further instructions. Before running these programs, the user first composes a specific search at the "Advanced Search" engine of the USPTO database of granted patents at http://patft.uspto.gov/netahtml/PTO/search-adv.htm. The routines work equally well with the sister database of patent applications at http://appft.uspto.gov/netahtml/PTO/search-adv.html. As noted, patent applications cannot be cited and, therefore, one can visualize patent applications only in terms of geographical concentrations and not in terms of citations as a proxy of the knowledge dynamics.

Using these databases, one can, for example, search with countries, states, or city addresses in addition to the issue and/or application dates of the patents under study. (The patent examination process may take years.) Instead of inventors, one may wish to delimit the set in terms of assignees and/or patent classifications, etc. However, the resulting recall must contain *at least 50 patents* so that one can move beyond the first 50 hits to the URL indicated for the 51$^{st}$ or any higher number (by clicking on "Next 50 hits" and then opening one or another patent in the browser).

The routine (uspto2.exe) begins by asking for this automatically generated search string—copied from the navigation bar of the browser for a patent with a sequence number higher than 50—and for the number of patents to be downloaded. Automatic downloading in the USPTO databases is limited to 1,000 at a time.[3] If more than 1,000 patents are wanted, the routine must be run more than once. The downloading of each 1,000 patents may take 15 minutes or so (depending on the quality of the connection).

The patents are automatically saved in the same folder on the hard disk as p1.htm, p2.htm, etc., in sequential order. Upon completion, one is first prompted with the question whether one wishes to map addresses of inventors or assignees. We focus here below on inventors. In the case of granted patents—with citations—another routine (patref3.exe) is automatically initiated that downloads the citation scores of the patents in files named sequentially q1.htm, q2.htm, etc. (These file names are used by the routines, and always overwritten by a next run. If one wishes to run thereafter for numbers higher than 1,000, the lower-numbered files should be saved elsewhere for combining them with the new downloads in a later stage.)

The various routines generate a series of output files that can be used for relational database management and statistical analysis. For this study, we use primarily the number of citations for each patent and the address information of inventors or assignees.[4] The full text of the citing patents would allow also to distinguish between applicant and examiner citations in US patents because the latter are asterisked on the front pages of the patents since 2001 (Alcácer *et al.*, 2009; cf. Criscuolo & Verspagen, 2008).[5] Without this extension—to be realized in the future—all citations on the front page of the patent are counted equally as a reference to prior art. Examiners add citations (both to patents and non-patent literature) if they consider these references as lacking given the legal status of patents.

---

[3] A notice with this warning can be found at http://www.uspto.gov/patft/help/notices.htm.

[4] In the case of patent applications, the addresses of assignees are not sufficiently standardized for reliable mapping onto Google Maps. However, the attribution of *assignee names* is reliable.

[5] The file "list2.txt" can be used for downloading the citing patents using the routine patref2.exe as explained at http://www.leydesdorff.net/lesson5.htm (Leydesdorff, 2004).

The output files "cit_inv.txt" and "cit_ass.txt," respectively, contain the addresses and these can be geo-coded with information about longitudes and latitudes. We used the facility of GPS Visualizer at http://www.gpsvisualizer.com/geocoder/ for this purpose, but one can also geo-code automatically, for example, using the (freeware) Sci2 Tool available at https://sci2.cns.iu.edu/user/download.php or the Stata commands developed by Ozimek & Miles (2011). The next routine (patref5.exe), however, needs a single input file named "geo.txt" using the format of GPS Visualizer, as follows:

```
latitude,longitude,name,desc,color
52.37312,4.893195,"Amsterdam, NL",-,
61.21759,-149.858354,"Anchorage AK, US",-,
etc.
```

If the file geo.txt is present, the user is prompted with further questions that influence the eventual layout of the map. First, the top-level of, for example, 10 or 25% has to be specified for the significance-testing. When mapping scientific publications, Bornmann & Leydesdorff (2011) used the top-10% of highly-cited papers as the default. However, publications in the Web of Science are an order of magnitude more numerous ($10^6$/year) than patents in the USPTO database ($10^5$/year). Therefore, the default is set in this case to the top-25% most-cited patents or the top-quartile. However, the user can change this.

For each city in the downloaded set the observed number of highly-cited patents is statistically tested against the expectation using the *z*-test for two independent proportions (Sheskin, 2011: 656). Without further *a priori* information, one can expect 1/4$^{th}$ of the patents from each city to belong to the first quartile. Statistical tests will only be performed for cities with expected values ≥ 5 because the *z*-test (like the chi-square) is not reliable for expected values lower than five. Significance levels are indicated (in the clickable descriptors of the cities on the map) as follows: * for $p < 0.05$; ** for $p < 0.01$; and *** for $p < 0.001$. Since maps may become overloaded with cities with one or a few patents (because of the possibly large tails of the distributions), the program suggests for the second question to assume a threshold of five or more patents/city, but this choice can be modified by the user.

This routine produces two output files (ztest.txt and patents.txt) which can be used as input to http://www.gpsvisualizer.com/map_input?form=data for generating the overlays. At this screen, one should choose the option "custom field" in the case of colorization and sizing: the label for custom colorization is "color," and "n" for size. The resulting map can be downloaded and edited as html. Furthermore, Google Maps distributes free APIs for uploading the map to one's website if so wished.

Using colors similar to those of traffic lights, cities with patent portfolios significantly below expectation in terms of citedness are colored dark-red and cities with portfolios significantly above expectation dark-green. Lighter colors (lime-green and red-orange) are used for cities with expected values smaller than five patents (which should not statistically be tested); light-green and orange are used for non-significant scores above or below expectation. The precise values are provided in the descriptors which can be made visible by clicking on the respective nodes. Additionally, all numerical values are stored in the database file "geo.dbf" for statistical analysis.[6]

Patents.txt contains the information for generating an additional map that is not based on citations, but on the portfolio of patents harvested with the search. Cities are here compared in terms of numbers of patents. This representation focuses more on geographical effects such as agglomeration and diffusion than on the knowledge dynamics. To that end, a quantile—that is, the continuous equivalent of a percentile—is counted for each city by dividing the number of cities with fewer patents (than the city under study) as the numerator, by the total number of cities in the set as the denominator. Using the same colors as Bornmann *et al.* (2011), the top-1% cities will be colored red (as "hot spots"), the top-5% fuchsia, the top-10% pink, the top-25% orange, the top-50% cyan, and the remainder (bottom-50%) is colored blue. These percentile rank classes follow the categorization used in the *Science and Engineering Indicators* series of the US National Science Board (2010; cf. Bornmann & Mutz, 2011).

The default maps are based on fractional (proportionate) counting of the addresses of inventors and assignees, respectively. In the case of some (industrial) laboratories, groups of inventors tend to sign together and thus a distortion is sometimes generated using whole-number or so-called "integer" counting. However, for the convenience of the user equivalent files called iztest.txt and ipatents.txt are also provided with the "*i*" for "*i*nteger counting" added.

---

[6] Differences between cities can also be *z*-tested for their significance as explained in Bornmann, De Moya-Anegón & Leydesdorff (in press). An Excel sheet available at http://www.leydesdorff.net/scimago11/index.htm can be used as guidance to this application.

## 6. Results

### *6.1. The Netherlands*

A search with "icn/nl and isd/2007$$" on October 3, 2011 recalled 1,908 patents issued in 2007 with a Dutch address among the inventors (icn = inventor's country name). Figure 1 shows the results as overlays to Google Maps (see for the interactive version at http://www.leydesdorff.net/patentmaps/nl_a.htm and http://www.leydesdorff.net/patentmaps/nl_b.htm, respectively ): 128 cities are found with five or more patents to their name.[7]

[7] In total 733 unique Dutch city names were found, but these included also misspellings. Misspellings are incidental in the USPTO database of issued patents.

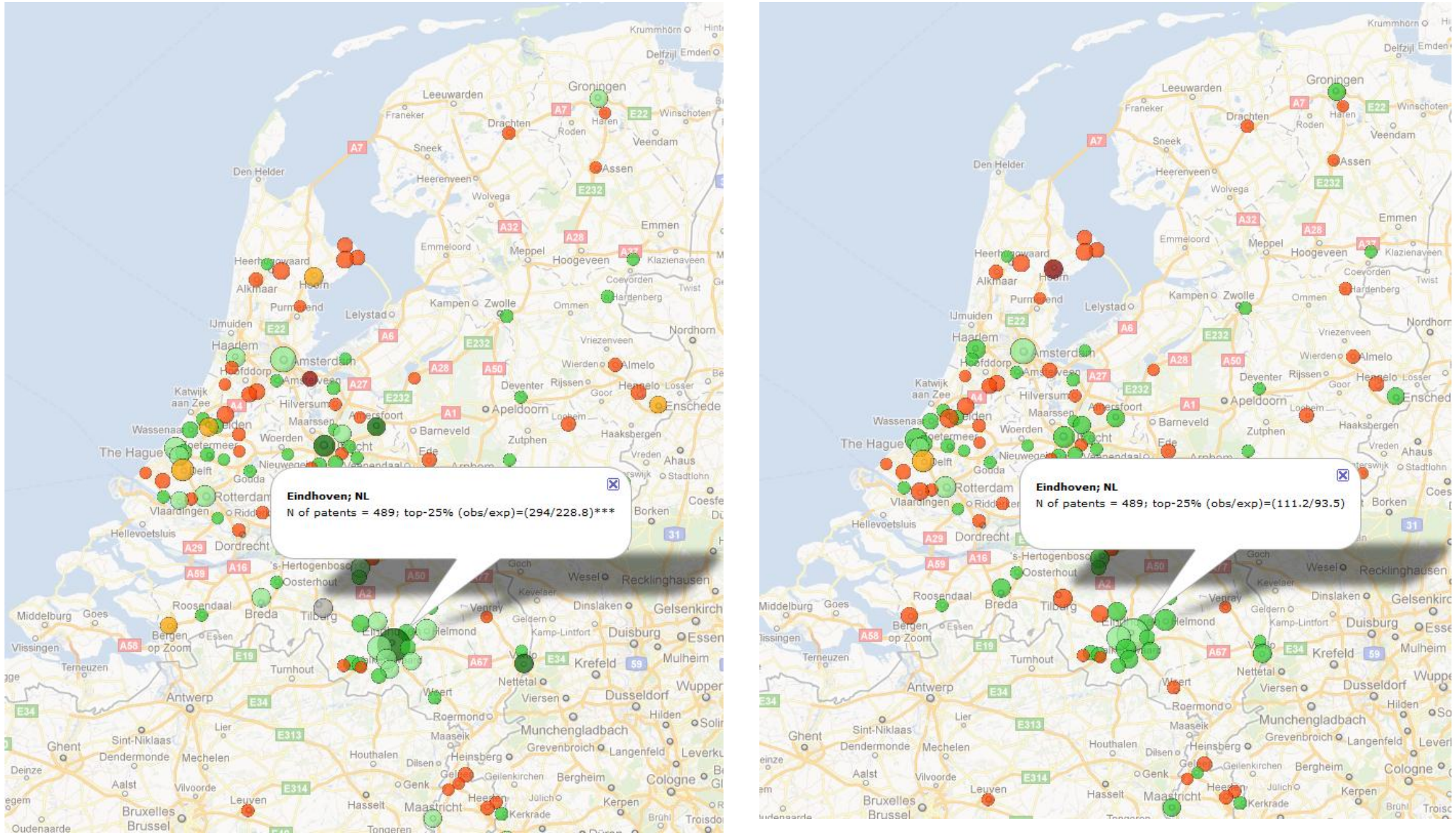


**Figure 1**: Portfolio for cities in the Netherlands with five or more patents, based on integer (left side) and fractional (right side) counting of the inventors. (See for interactive versions http://www.leydesdorff.net/patentmaps/nl_a.htm and http://www.leydesdorff.net/patentmaps/nl_b.htm, respectively.) The node sizes are proportionate to the logarithm of the number of patents.[8]

Among these cities, for example, "Eindhoven, NL" is in the address information of 489 patents with 915 inventors. In this data, each inventor or assignee is provided with a unique city name. The description window in Figure 1a (on the left side) shows that 294 of these inventors score in the top-quartile (top-25%) of the set in terms of patent citations. Using the *z*-test, this number is significantly higher than the (915/4 =) 228.8 patents expected in the top-quartile ($p < 0.01$), and therefore the node at Eindhoven is colored dark green. (The node sizes are proportionate to the logarithm of the number of patents.)[8]

---

[8] In order to prevent disappearance of a node in the case of maps including single patents, the *ln*(n+1) is used throughout this study.

However, 333 (68.1%) of these 489 patents are held by Philips Electronics in Eindhoven. As noted, patents in large laboratories are often signed by groups of researchers. Figure 1b (on the right side) corrects for this aggregation effect by fractionizing the number of inventors per patent, and providing every inventor (and thus every city) with a proportionate share. In this map, Eindhoven has the same number of 489 patents, but the score in terms of top-25% highly-cited patents is no longer statistically significant ($p > 0.05$). We use this normalization below as the default (but the routines also provide the results for integer counting).

Analogously, Weesp (a small town near Amsterdam) is colored dark red in the left-side figure because the number of top-cited patents is lower than expected. "Weesp" is in the address field of 44 inventors registered at the USPTO in 11 patents. Nine of these patents are held by the Solvay Laboratories; only one of these with five co-inventors[9] belongs to the top-25% group and this is statistically significantly below the integer-counted expectation of (44/4 =) 11 patents (Figure 1a). Using fractional counting, however, the number of inventors in Weesp is 4/5 in a single patent, against an expectation for the top-25% of 9.5/ 4 = 2.4. The node is colored orange because an expected value of 2.4 is too low for reliable testing.

[9] The patent names an American inventor in Bedford, MA, USA, as co-inventor. The co-assignee of this patent is Arqule Inc., MA, USA.

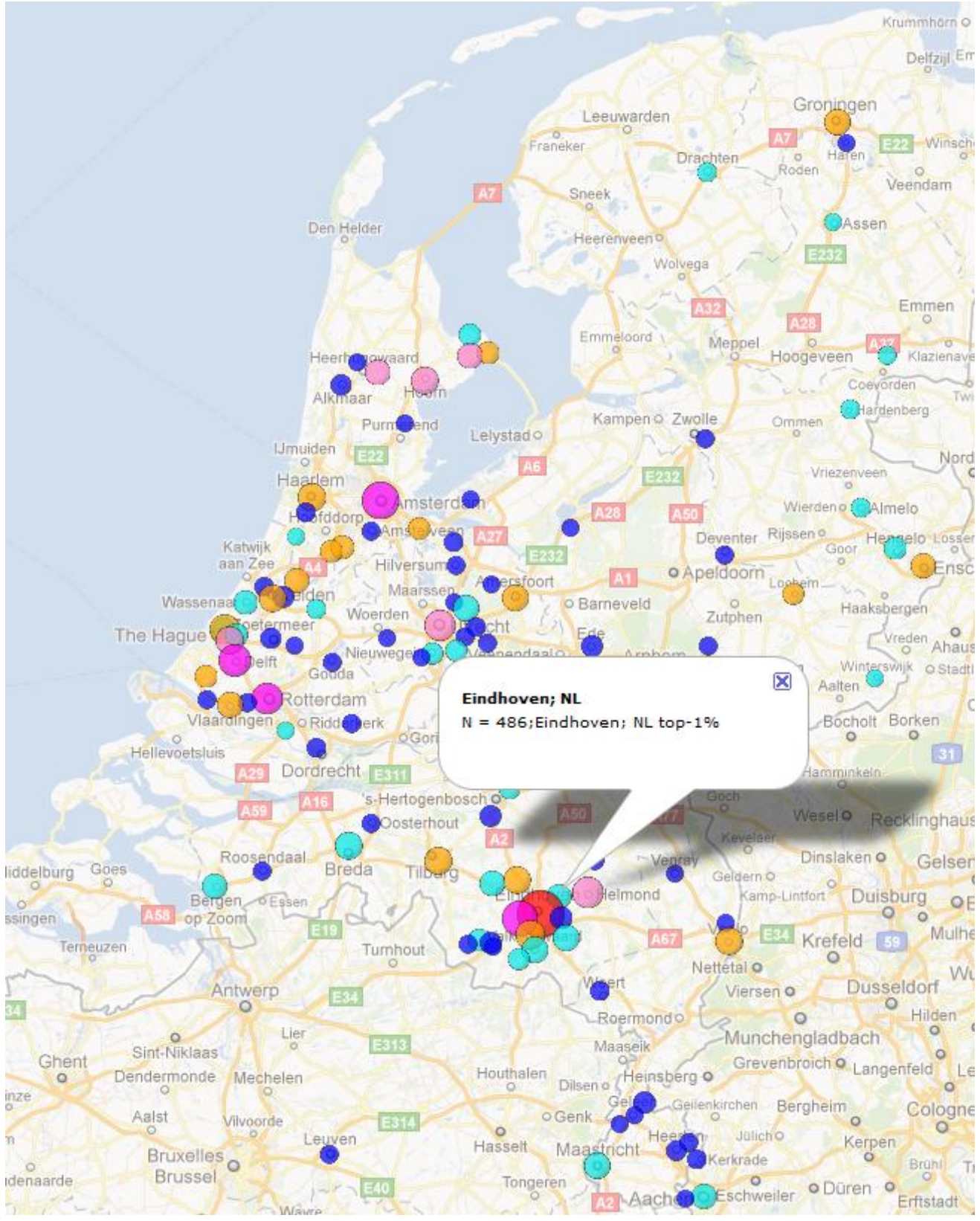


**Figure 2**: Portfolio for cities in the Netherlands with five or more patents in terms of percentiles: top-1% red; top-5% fuchsia; top-10% pink; top-25% orange; top-50% cyan; and bottom-50% blue. The node sizes are proportionate to the logarithm of the number of patents. For an interactive version see http://www.leydesdorff.net/patentmaps/nl_c.htm.

Figure 2 extends the analysis to patent portfolios as another visualization (based on patents.txt). This figure shows that "Eindhoven" with its 489 patents has a quantile value > 0.99 among 1,908 other Dutch patents. Amsterdam (112 patents), Delft (53), Rotterdam (44), Nijmegen (41), and Veldhoven (86; near Eindhoven and housing the ASML laboratories) follow at the top-5%-level in terms of numbers of patents.

These results accord with the perception of the most knowledge-intensive cities in the Netherlands, but the presence of Nijmegen—an older university city in the eastern part of the country—in the top-5% group was a bit of a surprise for us; Enschede with a

Technical University scored only in the top-25% category. Amazingly, Wageningen—supposedly the center of a cluster of innovations in the food sector (Porter, 2001, slide 43)—is visible in this representation with 12 patents only in the 56th percentile.[10]

These maps inform us that international patenting in the Netherlands is concentrated along the two highways A2 between Amsterdam and Eindhoven and A4 between Amsterdam and The Hague-Rotterdam. The former concentration is known in the Netherlands as more important than the latter (Van Oort & Raspe, 2005; cf. Draijer *et al.*, 2010; Leydesdorff *et al.*, 2006). In policy documents (e.g., Ministry of Economic Affairs, 2009: 4), the region Amsterdam-Utrecht is sometimes considered as distinct from the region Rotterdam-The Hague-Leiden (Otto Bernsen, *personal communication*, December 31, 2011). Using Google Maps, the overlay technique enables us to evaluate regional portfolios (Figure 2) and their quality in terms of patent citations (Figure 1b) by zooming in locally.

Zooming out at the Google Map, one can note that Houston, Texas—housing the assignee address of the Royal Dutch/Shell Oil—and Katy, Texas—housing a laboratory of Halliburton Energy Services—could also be considered as important centers in this set on the basis of co-inventions with 44 (top-25%) and 25 (top-50%) co-invented patents, respectively. Technology is not limited by national boundaries! We used the Netherlands above as a geographical unit for the exploration, but let us now consider the distribution of patents in the case of new or existing technologies.

[10] "Wageningen" is represented in the database with 15 inventors as co-inventors of 12 patents when integer-counted. However, fractionally counted these are only 4.69 patents. Both integer and fractionally counted these patents are rated as below expectation.

*6.2. RNA interference as an emerging technology*

In a recent study, Lundin (2011) analyzed patenting in the emerging technology of RNA interference (RNAi); in a previous study, one of us analyzed the diffusion of the research front in terms of the geographical locations of scientific publications (Leydesdorff & Rafols, 2011; cf. http://www.leydesdorff.net/et/sirna.html). With this background information as a kind of baseline, we chose RNAi as a first test case to map patent data for a specific research front.

"RNA interference" originated as a research program in molecular biology with an article published in *Nature* in 1998 entitled "Potent and specific genetic interference by double-stranded RNA in *Caenorhabditis elegans*" (Fire *et al*., 1998). The two principal investigators—Craig Mello at the University of Massachusetts Medical School in Worcester, and Andrew Fire at Stanford University—received the Nobel Prize in medicine for this breakthrough in 2006. RNA interference allows for control of the gene expression of DNA by inhibiting the messenger RNA. The new research program focused initially on the production of exogenous small RNA molecules with potentially large (e.g., therapeutic) effects (Sung & Hopkins, 2006). These molecules have become known as siRNA (that is, "small-interfering RNA"). More recently, the program was further extended to the micro scale: "micro RNA" (miRNA) generates the same mechanism endogenously within the cell.

In our previous study, a search with "(ts=siRNA or ts=RNAi or ts="RNA interference" or ts="interference RNA")" in the *Science Citation Index-Expanded* at the Web-of-Science (WoS) yielded a recall of 27,946 documents on February 10, 2010. At this same date, the recall at the USPTO was only 134 with a maximum value of 82 in 2009 when using the same search string on titles and abstract words. Using expert advice and further interview materials (Ismael Rafols, *personal communication*, October 4, 2011) the search string was extended with the newer concept of miRNA as follows: "TTL/(((siRNA$ OR RNAi$) OR microRNA$) OR miRNA$) OR ABST/(((siRNA$ OR RNAi$) OR microRNA$) OR miRNA$)". The recall for this string was 451 patents on October 7, 2011 (with more than 87% since 2009). Searching with the same string in the USPTO database of patent *applications* recalled 2,343 applications on October 14, 2011 (cf. Lundin, 2011).[11] We return to this larger set below, but let us focus first on the granted patents.

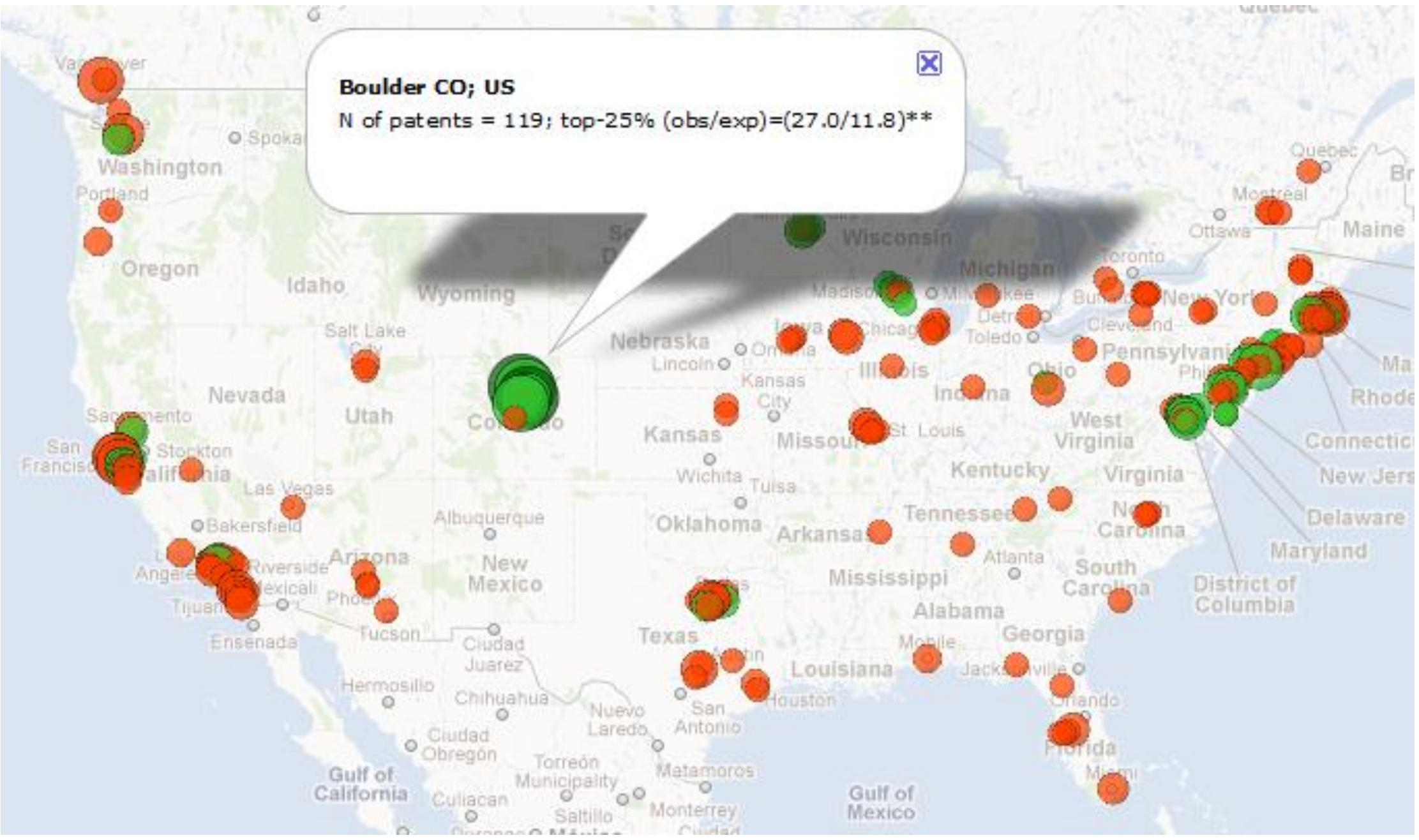


---

11 Our numbers are a bit lower than those of Lundin (2011) who used a more sophisticated search strategy, including patent classifications.

**Figure 3**: Quartiles (in terms of times cited) of patents in small and micro-interference RNA (2008-2010) in the USA; fractional counting. The interactive Google Map can be found at http://www.leydesdorff.net/patentmaps/sirna.htm .

The resulting Figure 3 shows that among the 412 cities with USPTO patents in this domain worldwide only patents in Boulder and Denver, Colorado, are colored dark-green, indicating significance levels of the top-cited patents above expectation ($p < 0.05$); 83 of the 451 patents are owned by *Dharmacon RNAi Technologies* in Lafayette, Colorado. These patents account for 960 of the 1,232 citations (77.9%). One other patent (16 times cited) is assigned to *SiRNA Therapeutics*, currently a research laboratory of *Merck*, but also in Boulder, Colorado. Various co-inventors of these patents provide home addresses in nearby cities in Colorado.

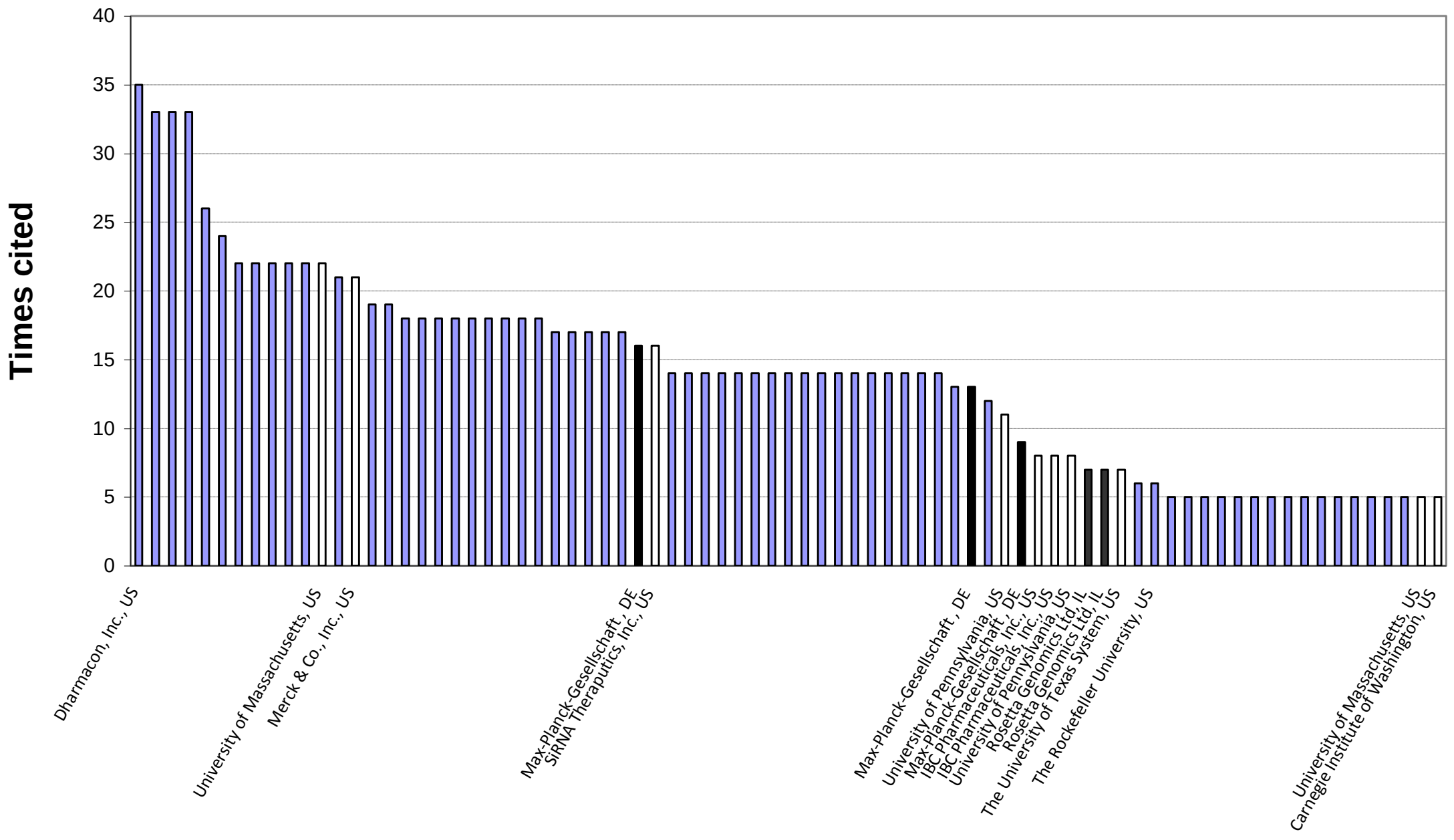


**Figure 4**: 79 patents in RNA interference (2008-2010) with five or more citations sequentially ordered according to the citation rate (October 2011).

Figure 4 shows the patent distribution for the 91 patents in the set that have received at least five citations in the meantime. The white bars indicate patents held by American assignees other than *Dharmacon Inc*., and the black ones non-US assignees. A "home advantage" effect can be expected using the USPTO (Criscuolo, 2006), but the map (showing all patents) reveals that most non-US cities are scoring at levels below expectation. The exceptions are two patents held by the German Max Planck Gesellschaft that have been constitutive for the field (the so-called "Tuschl"-patents; cf. MIT Technology Licensing Office, 2006).

As Lundin (2011) noted, *Dharmacon,* unlike the other major players in this field, is a reagent supplier and not a pharmaceutical drug developer. Although he expects another boosts in patenting in the near future, drug development in this field seems to have stagnated more recently. Note that *SiRNA* was bought by Merck in 2006 for the very high price of US$ 1.1 billion.

A citation analyst may further note that the citation curve of Figure 4 is not as skewed as one would expect in the case of scientific publications. Different from the latter, patents do not compete for citations. Inventions have been considered as non-rival in nature (Arrow, 1962; Romer, 1990). However, corporations compete in terms of patent portfolios. Aggregation of patents and citations for assignees shows the overwhelming dominance of *Dharmacon RNAi Technologies* (Lafayette, CO) in patenting this new technology in the USA.

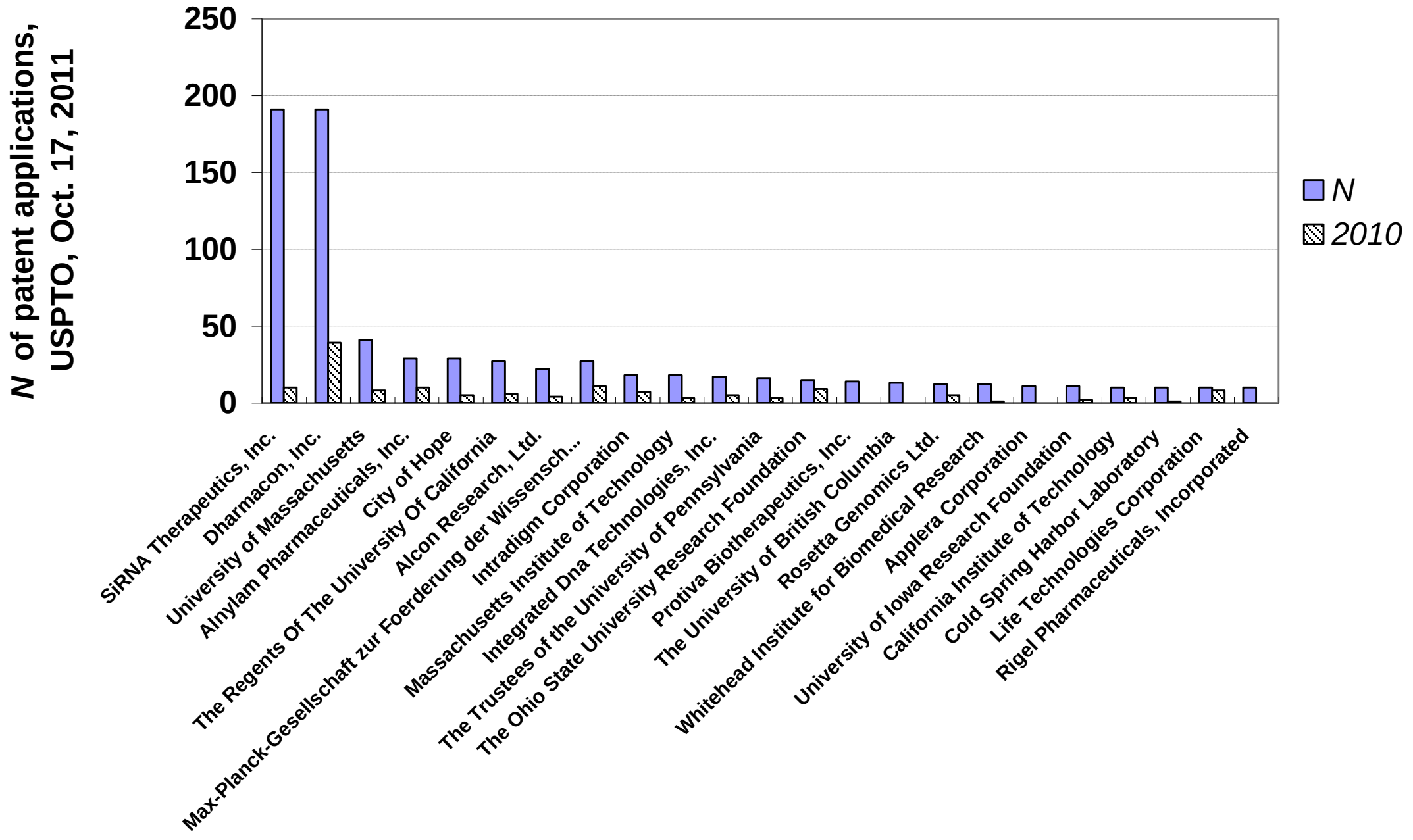


**Figure 5**: Assignees with ten or more applications among 2,343 patent applications with the USPTO since 2001; based on the database at http://appft.uspto.gov/netahtml/PTO/search-adv.html.

Figure 5 shows the frequency of patent *applications* in the US (cf. Lundin, 2011, Fig. 2b, at p. 494). In terms of these applications *SiRNA Therapeutics* and *Dharmacon* are equally sized. All other players are smaller in terms of numbers of applications. As noted, the advantage of using patent applications as indicators—instead of granted patents—is their closer proximity to the research front because the granting of patents often takes years. In order to focus on the last complete year (given the date of this research), we added the patent applications distribution in 2010 to Figure 5: 39 of the 364 applications in 2010 were filed by *Dharmacon*, followed by 10 applications each from *SiRNA Therapeutics* (San Francisco, CA) and *Alnylam Pharmaceuticals* (Cambridge, MA).[12]

[12] *Alnylam* was cofounded by one-time Max Planck researcher Tom Tuschl, Whitehead's Phillip Zamore and Phillip Sharp, and MIT's David Bartel who co-invented the constitutive Tuschl-I patent of 2001.

The map at http://www.leydesdorff.net/patentmaps/applications.htm shows the addresses of 8,439 (co-)inventors named in the 2,343 patent *applications*, colored in accordance with the six percentile ranks mentioned above. Red colored (top-1%) locations are: Austin (TX), Boulder, Conifer, Denver, Lafayette, Longmont (CO), San Francisco (CA), Brookline and Cambridge (MA), New York (NY), Rehovot in Israel, Vancouver in Canada, and Tokyo in Japan. European and Korean cities (Seoul) follow in the top-5% group. The strong concentration of patent applications in Colorado accords with the results for the granted patents reported above.

### *6.3. Nano technology*

Given the limitations in the above case of a single and emerging technology, let us turn to a field of patents more broadly defined, nanotechnology. The delineation of the nano-field itself is a subject of research, but refining the search string is not the subject of this study (Bonaccorsi & Vargas, 2010; Huang, 2011). Both the USPTO and EPO have created new (cross-reference) classification categories for nanotechnology—Class 977 and Y01N, respectively—but these classes do not (yet) function for searching at the user interface of the USPTO databases (Scheu *et al.*, 2006; cf. Leydesdorff, 2008).

The use of a search string has the advantage of avoiding an indexer effect in the classifications, but the recall remains necessarily incomplete.[13] Given our objective, we chose the rather basic string of "TTL/nano$ AND ((ISD/2008$$ OR ISD/2009$$) OR

[13] In the databases of Thomson-Reuters, such as the Derwent Innovation Index and the Science Citation Index, indexing is additionally tagged for the retrieval. Topic Searches (TS) include title words, abstract words, author keywords, and keywords additionally attributed by the indexing service.

ISD/2010$$)" and retrieved 2,947 patents on October 9, 2011. Including abstracts, one can retrieve 2,064 additional patents using this same string, but a sample of approximately 3,000 (that is, the data gathered by title search) is sufficiently large for the purpose of this study.

The first part of our search string guarantees that all words that begin with "nano" (e.g., "nanotubes") in the title will be included. By choosing the three last years, the information is recent yet older patents are given time enough to become cited. The expectation is that specific patterns of variation such as skewness in the distribution can be developed in a larger set and over time (Barabási & Albert, 1999; Leydesdorff & Rafols, 2011; Price, 1965).

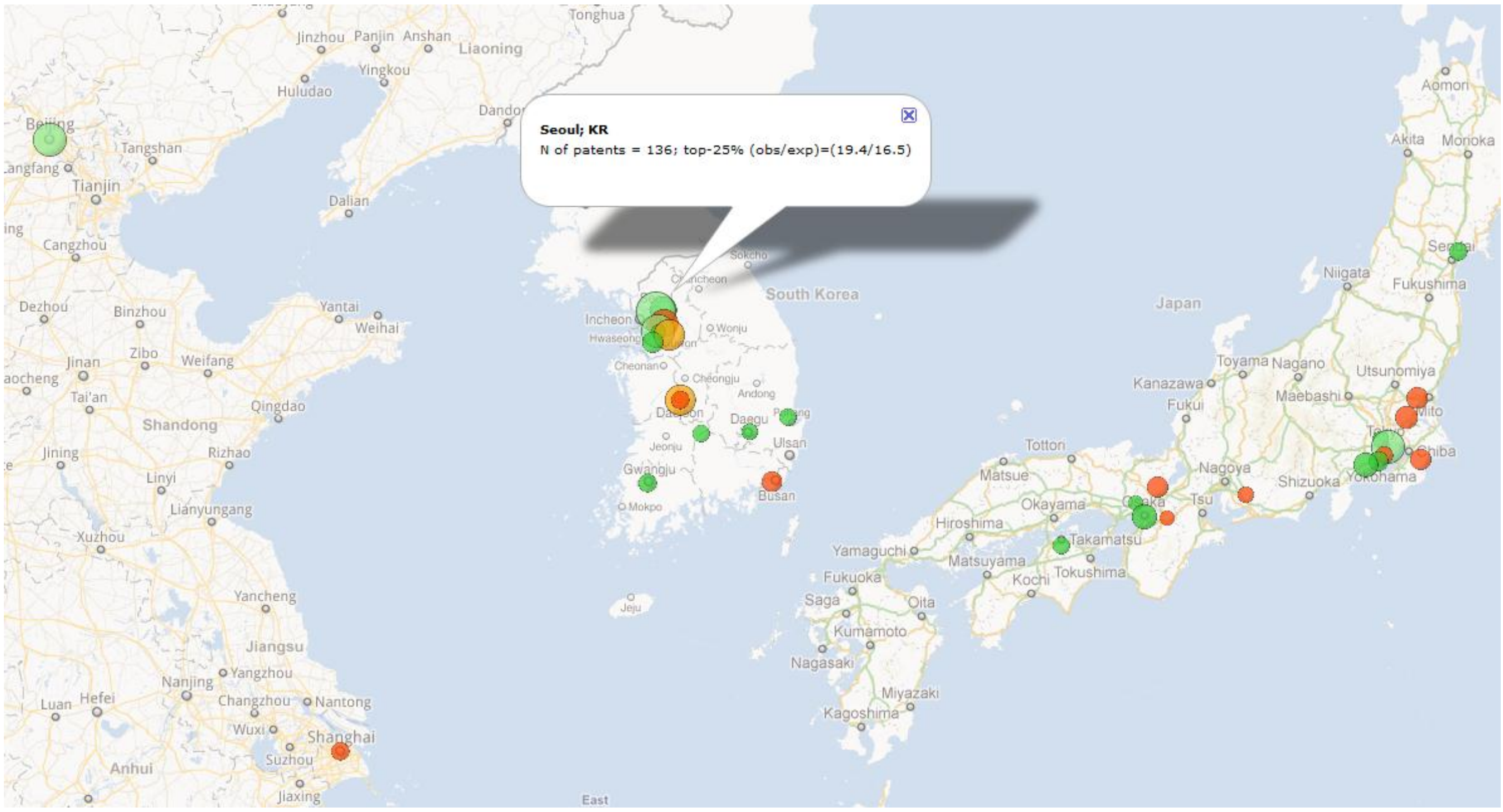

**Figure 6**: Fractionally counted distribution of patents with "nano*" in the title (2008-2010); $N$ of patents $\geq 5$; fractional counting; ($z$-)tested against the expectation of 25%

most-highly cited patents. The interactive map is available at http://www.leydesdorff.net/patentmaps/nano_b.htm.

The maps constructed on the basis of this search string using integer and fractional counting, respectively, can be found at http://www.leydesdorff.net/patentmaps/nano_a.htm and http://www.leydesdorff.net/patentmaps/nano_b.htm. Both Seoul and a number of American cities score significantly above expectation when integer counting is used. For example, 136 patents are attributed to Seoul with 234 co-inventors. Fractionally counted, however, only 65.8 patents are co-invented in Seoul because of the large teams of co-inventors from other (Korean) locations. The 19.4 patents observed in the top-quartile in this case are tested against (65.8 / 4 =) 16.5 expected; this difference is no longer statistically significant (see Figure 6).

Figure 6 also shows that the east coast of mainland China is virtually absent from the map (given a threshold of five patents per city in this case). In general, China has low numbers of patent applications to the USPTO (Shelton & Leydesdorff, 2011). However, Western Europe has also few USPTO patents in this domain. This impression can further be supported by coloring the percentiles using the entire patent distributions (at http://www.leydesdorff.net/patentmaps/nano_c.htm). With the exception of Malmö (Sweden; 11 patents) and Dresden (Germany; 9 patents) in the top-5% of the distribution, the cities in Europe are mostly colored blue indicating a modest contribution (bottom-50%). (Note that this analysis is based on fractional counting.)

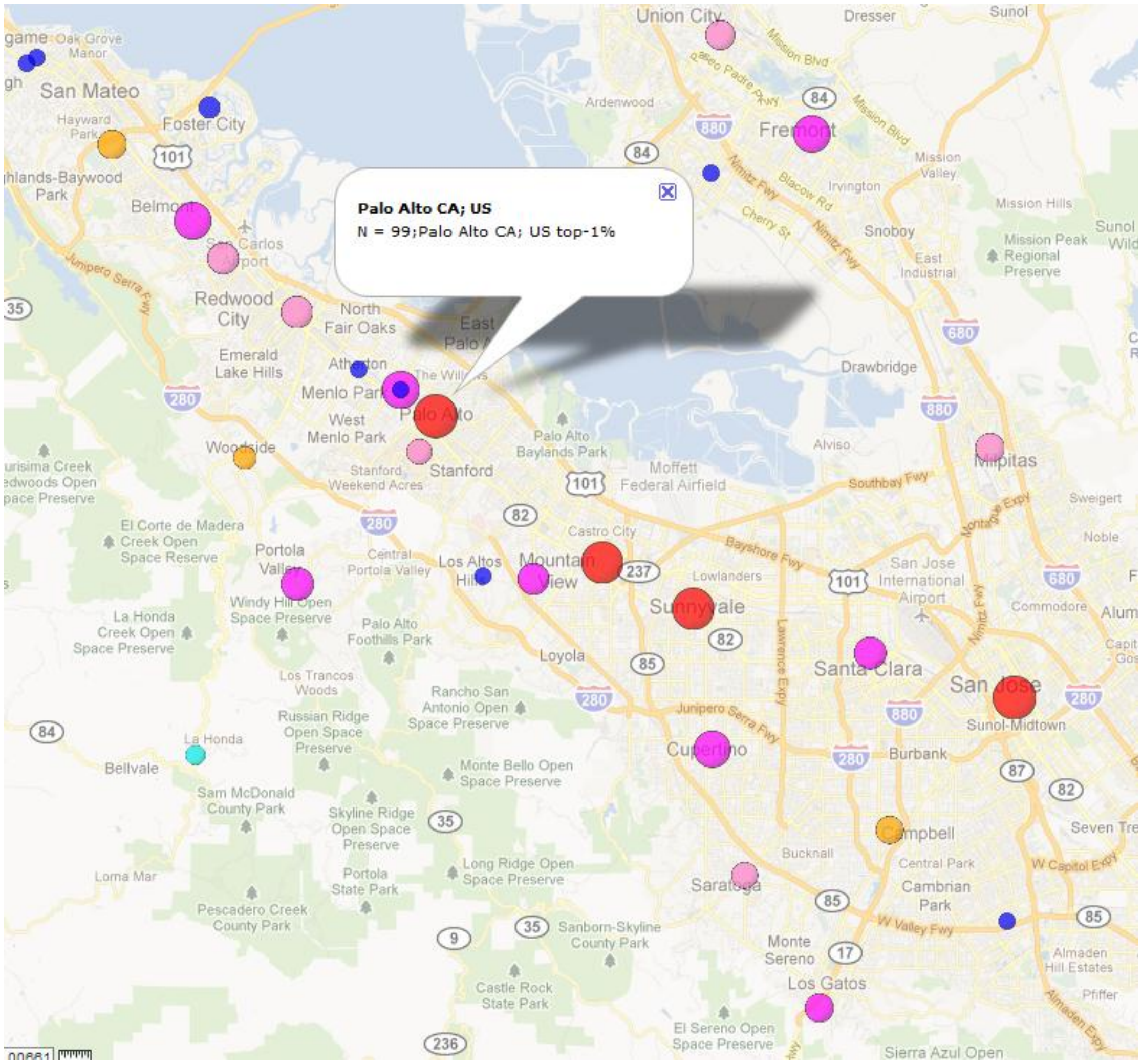

**Figure 7:** Distribution of the numbers of patents in Silicon Valley (California, US); *N* of patents = 2,964; *N* of cities = 2,090; fractionally counted attributions. The interactive map can be found at http://www.leydesdorff.net/patentmaps/nano_c.htm .

Figure 7 shows the high concentration of patents in Silicon Valley. Palo Alto (Stanford) takes the lead with 99 patents, but San Jose (98), Sunnyvale (78), and Mountain View (77) also score in the top-1% of the (fractionally counted) patent distribution. A similar high concentration can be found in the metropolitan area of Seoul, and to a lesser extent in southern Texas (Austin and Houston). Other regions of academic and entrepreneurial activity such as the Boston area are also indicated.

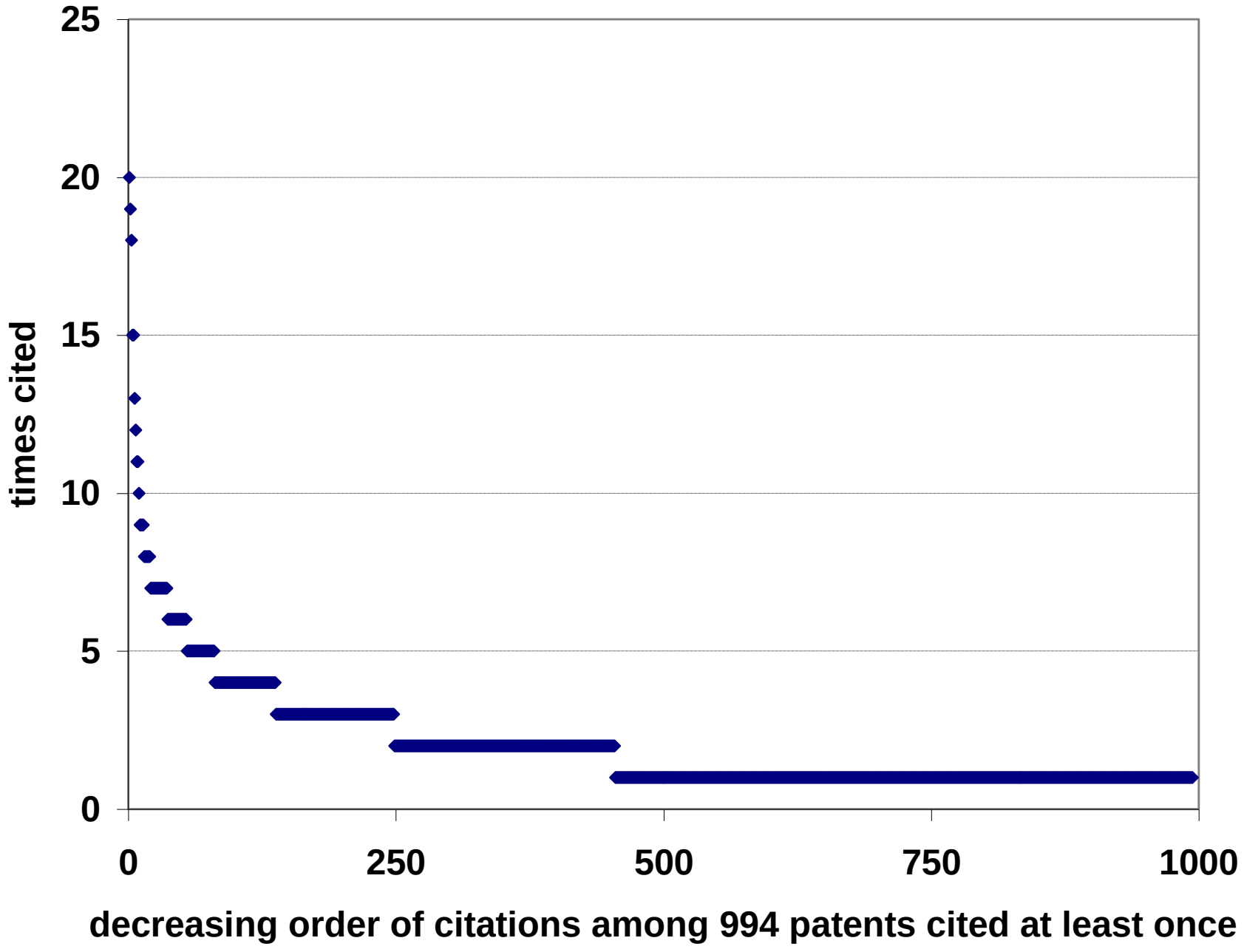

**Figure 8**: Citation curve of 994 (of the 2,947) "nano-"patents cited at least once.

Figure 8 shows the skewed pattern in the citation curve of these patents, but this citation curve is perhaps less extreme in the top values than in the case of citation distributions among journal publications (e.g., Katz, 2000; Leydesdorff & Bornmann, 2011). This observation confirms the previous impression that citation distributions of patents are less skewed than for scholarly literature. As noted, the competition mechanism is very different in these two domains.

| | Times cited |
|---|---|
| 3M Innovative Properties Company, St. Paul MN, US | 20 |
| QuNano AB, Lund, SE | 19 |
| The Goodyear Tire & Rubber Company, Akron OH, US | 15 |
| The Goodyear Tire & Rubber Company, Akron OH, US | 15 |
| HRL Laboratories, LLC, Malibu CA, US | 13 |
| 3M Innovative Properties Company, St. Paul MN, US | 12 |
| Spire Corporation, Bedford MA, US | 11 |
| QuNano AB, Lund, SE | 11 |
| President and Fellows of Harvard College, Cambridge MA, US | 10 |

**Table 1:** Assignees of the nine most-cited patents (cited ten or more times) among 2,947 patents in "nano-technology".

The assignees of the nine most-cited patents in the set (cited ten or more times) are listed in Table 1. Three companies own more than a single patent in this subset, and one of these is not from the US, but Sweden. Another of these patents is held by a university (Harvard). Thus, the pattern is more distributed than in the previous case.

## 7. Conclusions and discussion

As mainly descriptive techniques—despite the *z*-testing—the explorations above lead primarily to results raising questions and giving directions for further research. However, the routines produce also the file geo.dbf which contains all the quantitative information generated. This data can be used for various (hypothesis-)testing purposes.

For example, the strong position of one corporation and one specific region in the American patent market in the case of RNA interference came as a surprise. Previous research focusing on publication patterns pointed in the direction of an oligopolistic structure among centers of excellence in metropolitan areas (such as Boston, London, and Seoul) with perhaps a division of labor among them (Leydesdorff & Rafols, 2011). Using patent applications and other databases, Lundin (2011) signaled a concentration of patenting and the corresponding shift of industrial emphasis in the exploitation, but may have underestimated, in our opinion, the statistical significance of it.

Industrial competition on the knowledge market of patenting is structured very differently from the competition among scholars for publications and citations as quality indicators. The differences in success rates for different corporations (or perhaps even industries), for example, were highly significant in the case of RNA interference. Patenting is more competitive than publishing because higher costs are involved and the number of patents in the USPTO is restricted because of the examination procedures in place. The differences can be by orders of magnitude.

It came as a surprise to us that the citation distributions of these patents were relatively flat when compared with publications. Therefore, we lowered the (default) threshold to the top-25% (instead of the top-10%) as indicators of excellent impact. The use of these seemingly arbitrary thresholds can technically be overcome by testing for percentile-normalized citation distributions, as in the case of the Integrated Impact Indicator (*I3*) that we recently proposed in the context of citation analysis of the journal literature (Leydesdorff & Bornmann, 2011).

Following the recent introduction of the top-10% as excellence indicators in the *SCImago Institutions Rankings 2011* and the *Leiden Rankings 2011/2012* of 500 top-universities, we suggest to distinguish between impact and excellence indicators. *I3* can be considered as an impact indicator (cf. Prathap, in press), but the top-10% or the top-quartile—as in this study—provide us with an excellence indicator (Bornmann *et al.*, in press). Alternatively, one can adopt the relatively straightforward six percentile ranks (top-1%, top-5%, top-10%, top-25%, top-50%, and bottom-50%) developed by the US National Science Board (2010) in *Science and Engineering Indicators* (Bornmann & Mutz, 2011; Leydesdorff *et al.*, 2011).

Our primary objective was to make the technique available as an "instrumentality." The term "instrumentality" was introduced by Price (1984: 13) as a general connotation of a method for doing something to the data at hand. Instrumentalities serve the opening of new domains of research. In this case, the overlays provide us with "patent radar" so that one can signal unexpected information. This may lead to new research questions such as in the case of the Netherlands, the more pronounced function of the highway Amsterdam-The Hague-Rotterdam when compared with Amsterdam-Utrecht-Eindhoven; in the case of RNA interferences the already signaled shift from the exploration of new drugs to the exploitation of reagents; and in the case of nanotechnology, the relative minor role of cities outside the USA (with the exception of Seoul in Korea) for patenting in this high-tech market. The position of Silicon Valley in this latter case was expected, but confirmed at a level above expectation.

A longer-term objective is the transversal combination of information across databases such as the WoS or Scopus, patent databases, and perhaps PubMed in order to trace innovation trajectories from an information-science perspective (cf. Swanson, 1990; Fagerberg & Verspagen, 2009). The geographical combination of (address) information across the various databases can be integrated into overlays to Google Maps. The current project added this step for the USPTO data. More importantly, we envisage combining demand-side and supply-side information at the cognitive level in future projects (De Nooy & Leydesdorff, in preparation). A third dimension could be the mapping of corporate structures in relation to these two dimensions of substantive and geographical information flows using the Triple Helix model (Leydesdorff *et al.*, 2006).

An issue in the case of such an envisaged integration of maps remains the quality of each instrument and each database. As noted, the misspellings were more numerous in the database of patent applications than in granted patents. Of course, our routines cannot correct for misspellings in the basic materials of the USPTO databases, but on a case-to-case basis we were able to correct some errors. For example, in the current routines we correct for the confusion caused by using, for example, both "Gyeonggi-do" and "Gyeonggi-Do" in South-Korean patent addresses. Such systemic errors can be corrected at the file level when they are found, and we are eager to receive feedback from colleagues whenever possible. (In the case of analyzing patent *applications*, the output data is less reliable because of misspellings in the address field.)

Further extensions can be envisaged for using other databases with open access such as the search interface of the World Intellectual Property Organization (WIPO) in Geneva at http://www.wipo.int/patentscope/search/en/advancedSearch.jsf which houses the patents under the Patent Cooperation Treaty (PCT). This database, however, does not contain patent citations. As noted, the European Patent Office (EPO) provides a sophisticated interface at http://worldwide.espacenet.com/advancedSearch?locale=en_EP that couples to 80+ patent offices worldwide, but downloading sets of more than 500 is difficult. A more dedicated search engine can be found at the European Publication Server (https://data.epo.org/publication-server/?lg=en), but with limited search facilities.

In addition to its transparency, the USPTO database has the advantage of being the prime indicator of new technological inventions, and is therefore the most relevant one for innovation policies (Narin & Olivastro, 1988; Jaffe & Trajtenberg, 2002). As noted, our programs can function with both the database for granted patents and the one for patent applications. The latter follow the research front more closely than the granted patents, but do not contain citation information. More recently, foreign grants for USPTO patents outnumbered American grants (Shelton & Leydesdorff, 2011) thus moderating the "home advantage" effect of this database.[14] The two excellent search engines of the USPTO allows the user to generate specific sets in great detail, and as indicated, the barrier of 1,000 patents at a time can be circumvented by using different starting numbers in the various routines.

**Acknowledgement**
We are grateful to Ismael Rafols, Duane Shelton, and Fred Ye for comments on previous drafts. The first author acknowledges support by the ESRC project 'Mapping the Dynamics of Emergent Technologies' (RES-360-25-0076).

[14] Foreign applications have outnumbered American ones for years.